\documentclass[12pt]{article}

\usepackage[version=3]{mhchem}      
\usepackage{epstopdf}
\usepackage[super,sort&compress]{natbib}
\usepackage[pagewise,modulo]{lineno}
\usepackage{graphicx}
\usepackage[affil-it]{authblk}
\usepackage{multirow}
\usepackage{chemscheme}
\usepackage{amssymb}
\usepackage{mathtools}

\usepackage{xcolor}
\usepackage{algorithmic}
\usepackage{algorithm}
\usepackage{bm}

\usepackage{dsfont}

\author[1]{Chiara Faccio}
\author[1]{Michele Benzi}
\author[2]{Laura Zanetti-Polzi\thanks{email: isabella.daidone@univaq.it; laura.zanettipolzi@nano.cnr.it}}
\author[3]{Isabella Daidone$^\ast$}

\affil[1]{Scuola Normale Superiore, Piazza dei Cavalieri 7, 56126 Pisa, Italy}
\affil[2]{Center S3, CNR-Institute of Nanoscience, Via Campi 213/A, 41125 Modena, Italy}
\affil[3]{Department of Physical and Chemical Sciences, University of L’Aquila, via Vetoio (Coppito 1), 67010 L’Aquila, Italy}

\date{}

\title
{Low, high and very-high density forms of liquid water revealed by a medium-range order descriptor}

\begin{document}
\maketitle

\begin{abstract}

We present in this paper a computational approach based on molecular dynamics simulations and graph theory to characterize the structure of liquid water considering not only the local structural arrangement within the first (or second) hydration shell, but also the medium- to long-range order. 
In particular, a new order parameter borrowed from the graph-theory framework, i.e. the node total communicability (\emph{NTC}), is introduced to analyze the dynamic network of water molecules in the liquid phase. 
This order parameter is able not only to accurately report on the different high-density-liquid (HDL) and low-density-liquid (LDL) water phases postulated in the liquid–liquid phase transition hypothesis, but also to unveil the presence of very high density liquid (VHDL) clusters, both under pressure and at ambient conditions. To the best of our knowledge, VHDL water patches under moderate pressures were not observed before.

\end{abstract}

\section*{Introduction}

The structural characterization of liquids (and glasses) is far more difficult than that of 
 crystalline solids because, differently from the latter, liquids lack long-range crystalline order, while retaining short-  to medium-range order. This difficulty is particularly relevant when discussing anomalous kinetic and thermodynamic properties, such as those of liquid water. One of the most popular hypotheses for explaining many of the water anomalies is based on the existence of two metastable liquid forms of water with different densities, referred to as low-density liquid (LDL) and high-density liquid (HDL), respectively.\cite{poole1992phase} In particular, computational studies have proposed a picture of a first-order liquid–liquid phase transition (LLPT) between these two metastable liquids ending up at a metastable critical point.\cite{poole1992phase,gallo2016water,palmer2018advances,debenedetti2020second,Gartner26040,zanetti21,sciorts21JCP} 

The two different HDL and LDL liquids have their counterparts in the glassy state, that is, the high-density amorphous (HDA) and low-density amorphous (LDA) forms of ice.\cite{Mishima84,Loerting06}
A third form of amorphous ice, a very high-density amorphous (VHDA) form, was also observed and shown to be distinct from HDA.\cite{loerting2001second} 
Neutron-scattering data showed a shift of the second peak in the O$\cdots$O-pair distribution function, $g_{OO}(r)$, from $\approx$0.45 nm in LDA to $\approx$0.40 nm in HDA and $\approx$ 0.31-0.35 nm in VHDA. This shift corresponds to an increased population of interstitial water molecules that, in the high and very high density forms, move closer to the first coordination shell.\cite{fuentes15} It was recently suggested that this increase might arise from the "folding" of water rings which brings molecules separated by three or four hydrogen-bonds close in space. \cite{sciorts21PRL} 

It has been hypothesized that in systems with rich polymorphic solid phase diagrams, several liquid phases might also exist.\cite{brazhkin2002new}
The local particle structure in these liquid phases should resemble that of the corresponding solid structure. In line with this hypothesis, the existence of the liquid-counterpart of the VHDA ice (i.e. a very-high density liquid, VHDL) has been put forward by means of computational studies.\cite{paschek2008thermodynamic,buldyrev2003system}
 However, it is still an open question how exactly the different forms of amorphous ice and supercooled liquid water are connected, since the “no-man’s land” largely prohibits direct experimental access.\cite{debenedetti2003supercooled,giovambattista2005relation} 
Investigation of the VHDL phase has been addressed in a few computational works\cite{paschek2008thermodynamic,buldyrev2003system} and, to the best of our knowledge, the presence of VHDL was only observed at very high pressures/densities and was never observed as a spontaneously emerging phase at more moderate pressures or at ambient pressure.

Identification of a liquid phase in a simulation relies on the description of the structure of the liquid at the molecular level. The typical two-body correlation functions used to describe the structure of a fluid, the  structure factor and the radial distribution function, are however spherically averaged quantities that can not be used to characterize the local structure at the microscopic level. In addition, the limited variation of these functions in temperature often does not reflect the significant variation in the dynamic behavior of the supercooled fluid.\cite{tanaka20} 
The inclusion of many-body effects, as packing and excluded-volume effects, has been suggested to be crucial for a proper description of liquid states, especially in the supercooled regime.\cite{tanaka12,tanaka20} 
It is however not straightforward to obtain meaningful structural descriptors for disordered microscopic structural arrangements.

Here, we present a new descriptor for the structural characterization of liquids at the single-molecule level with the aim of taking into account medium-range order information and many-body effects. For this, we consider measures derived within the framework of graph theory. Indeed, graph theory provides several tools to analyze the structure of networks (in the present case the network of the oxygen atoms of the water molecules) and to understand the role played by each node (i.e. each water molecule). In this respect, of great importance are {\em centrality measures}, which are quantitative measures that aim at revealing the "importance" of each node within a network.\cite{BORGATTI200555,Borgatti2006AGP} 
Besides quantifying  the role played by a specific node in a network, centrality measures can also be used to characterize global properties, such as how well connected the network is overall, how easily different parts of the network communicate with one another, and how robust a network is against attacks aimed at disconnecting it.\cite{estrada2012structure,10.1093/comnet/cnt007}

In this paper we show how these easily computable centrality measures can be used  to characterize the networks of liquids (and in perspective of solutions).
In particular, we show their application to the problem of identifyng LDL/HDL phases from molecular dynamics trajectories and also to characterize possible clusters of one phase within the other.
We mainly focus on a specific centrality measure: the node total communicability. This measure accounts for the cumulative connectivity (from the short to the medium/long-range) of each network node. It is thus a descriptor of the packing of the hydration shell of each water molecule in the network.  
Although network based approaches have been applied to the investigation of other water-related processes \cite{bako2013hydrogen,karathanou2020graph,choi2021effects}, the problem of identifying the LDL and HDL phases of water using centrality measures has not been considered before.  
Moreover, the present approach was also able to bring to light the existence of clusters of the third, postulated, form of liquid water, i.e. the VHDL form.

\section*{Methods}

\subsection*{Review of basic graph theory and centrality measures}

We recall some basic notions and ideas from graph theory, see for example \citet{aa42cae0c7624473bb4216975da02d87}. 
A graph is a pair $G = (V, E)$, where $V = \{v_1,...,v_N \}$ is the set of nodes and $E \subset V \times V$ is the set of edges. A graph is directed if the edges have a direction, undirected otherwise.  In this discussion, we consider only undirected graphs. In other words,  $(v_i, v_j)$ and $(v_j, v_i)$ will denote one and the same edge. We do not consider self-loops, i.e., edges of the form $(v_i, v_i)$. 

The {\em adjacency matrix} associated with the graph  $G$ is a matrix $A = (a_{i,j})$ in which both the rows and columns are indexed by the vertices and $a_{i, j} \ne 0 $ if and only if there exists an edge from node $v_i$ to node $v_j$. We think of $a_{i,j}$ as the {\em weight} associated with the edge between $v_i$ and $v_j$. 
Note that $A$ is symmetric if and only if the graph is undirected. Here we assume that all nonzero weights are positive real numbers, so that $A$ is a real matrix with non-negative entries. If all edges in $E$ are given the same weight $a_{i,j}=1$, we say that the graph $G$ is {\em unweighted}. 
By the Perron-Frobenius theorem, the largest eigenvalue of $A$, denoted by $\rho (A)$, is real and positive. If the graph is strongly connected, meaning that there is a path connecting any two vertices in $G$, $\rho (A)$ is a simple eigenvalue and there exists a unique (up to normalization) eigenvector associated with $\rho(A)$ with strictly positive entries.
If the graph is unweighted, the $(i,j)$ entry of the matrix $A^k$ ($k$th power of $A$) is equal to the number of walks of length $k$ connecting $v_i$ and $v_j$, for all $k=0, 1, \ldots$.

For every node $v_i \in V$, the degree ($deg(v_i)$) is the sum of the weights of the edges incident upon the node $v_i$. 
If the graph is unweighted, the degree of a node is just the number of edges incident on it or, equivalently, the number of vertices adjacent to it. Note that $deg(v_i)$ is just the sum of the entries in the $i$th row of $A$. 

Next, we review some centrality measures.
The \textit{Degree Centrality} (\emph{deg}) is the simplest centrality measure and is the degree of a node. This measure considers only the first neighbours of a given node.

There are other measures that consider more general types of walks joining nodes in $G$. The \textit{Eigenvector Centrality} (EC)\cite{estrada2012structure,Bonacich1987PowerAC}  of the node $v_i$ depends on the importance of the neighbours of the node: a high centrality value means that the node is connected to nodes that have themselves a high centrality score.
 If the graph is strongly connected then, by the Perron-Frobenius Theorem, there  is (up to normalization) a unique positive eigenvector associated with $\rho(A)$, i.e., there exists an essentially unique vector $\bm{p} > \bm{0}$ such that $A \bm{p} = \rho(A) \bm{p}$. The eigenvector  centrality of node $v_i$ is defined as the $i$th entry of this eigenvector $\bm{p}$. It can be shown that
\begin{equation}
EC(v_i) = p_i = \lim_{k \to \infty} \frac{ \# \text{walks of length k through } v_i}{ \# \text{walks of length k in G}}
\end{equation}
This measure takes into account how well connected a node is and how many links their connections have, and so on through the entire network. 
It identifies nodes with influence over the whole network, not just those directly connected to it. 

A closely related notion is that of \textit{Total Communicability}\cite{10.1093/comnet/cnt007,ESTRADA201289} (per node), defined as follows: let $\beta > 0$, then
\begin{equation}
NTC(v_i) = [e^{\beta A} \mathds{1}]_i = \sum_{k = 0}^{\infty} \frac{\beta^k}{k!}[A^k \mathds{1}]_{i} 
\label{Eq:TC}
\end{equation}
where $\mathds{1}$ is the vector of all ones. This measure considers all walks between node $v_i$ and every node in the network
but it gives less weight to longer  walks. This  is achieved through the use of  the parameter $\frac{\beta^k}{k!}$, where $\beta$ can be tuned to give more or less weight to longer  walks; for large $\beta$, this measure is equivalent to Eigenvector Centrality, for small $\beta$ to Degree Centrality \cite{Benzi2015OnTL}. Here $\beta$ is chosen to be equal to 1. A distinct advantage of the total communicability is that it can be computed much more efficiently than other centrality measures in the case of large networks.

In Scheme \ref{scheme1} the different ranking of the nodes provided by the degree \emph{deg} and the node total communicability \emph{NTC} is represented for a very simple network.

\begin{scheme}
\centerline{
\resizebox*{12.00cm}{!}{\includegraphics{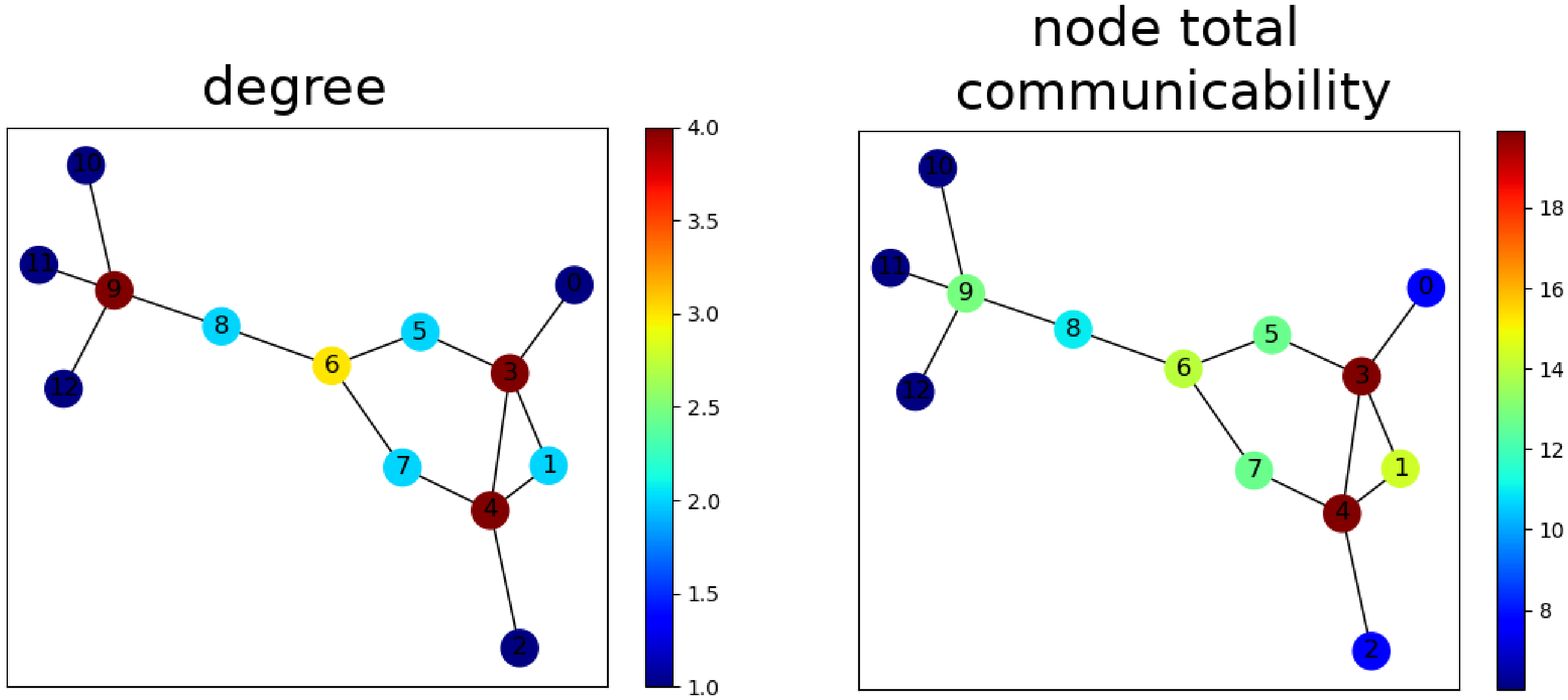}}}
\caption{Representation of the ranking provided by the degree \emph{deg} and the node total communicability $NTC$ for a very simple network. The nodes are labeled from 0 to 12, and they are colored according to their centrality measure: red nodes have the highest centrality measure, blue nodes have the smallest value of centrality in the network.}
\label{scheme1}
\end{scheme}

\subsubsection*{Construction of the adjacency matrix}

In this work we aim to analyze water molecules in a box. To this end, we build a graph $G = (V, E) $, where each water molecule corresponds to a vertex $v_i\in V$ and a bond between two water molecules is represented by the edge $e_{i j}\in E$. To identify bonds, we use the following criteria: two molecules of water are bonded when the distance between oxygen atoms is $\le 0.35$ nm. In this way, the corresponding graph is undirected and the adjacency matrix is symmetric of dimension $N \times N$, where $N$ is the number of water molecules in the box. In future works, we will also consider the arrangement of the protons. 

The water molecules move in time according to the laws of Molecular Dynamics (MD). 
Given a MD trajectory, at each time step, we extract a file %(named file.gro) 
with the coordinates of the atoms. Since in the simulation we impose periodic boundary conditions, meaning that the box is replicated along the three directions of space, each molecule interacts with the images of the molecules which are on the opposite sides of the box. To ensure that the calculated interactions of the box are more realistic, we also consider periodic images in the analysis. In this first work, we do not consider the angles, so we replicate the box along the x, y, and z axes. In this way, we have seven boxes: six replicas and a central box. With this new set of coordinates, we construct the adjacency matrix as explained before and calculate the centrality measures of its nodes. At the end of the analysis, we consider the values of the middle box only.

All network analyses are carried out using the NetworkX 2.5.1 module \cite{SciPyProceedings_11} in Python 3.7 and the Python package \emph{sobigdatanight}.\cite{Bucci}

\subsection*{Molecular dynamics simulations}

We use here previously performed\cite{zanetti21} 150 ns-long MD simulations of neat water at 1950 bars at four temperatures (170 K, 180 K, 200 K and 240 K) and a 50 ns-long trajectory at 1 bar and 300 K. For all MD simulations, the TIP4P/2005 water model\cite{abascal2005general} was used, as it was shown to exhibit a metastable liquid–liquid critical point under deeply supercooled conditions (at 1700-1750 bar and 177-182 K depending on the details of the simulaiton conditions).\cite{debenedetti2003supercooled,abascal2010widom,sumi2013effects,biddle16,biddle2017two} MD simulations were performed with the GROMACS package (version 5.1.2)\cite{abraham2015gromacs} in the NPT ensemble using a rectangular simulation box, the velocity rescaling temperature coupling,\cite{bussi2007canonical} and the Parrinello–Rahman barostat with 2 ps relaxation times.\cite{parrinello1980crystal} Periodic boundary conditions were used, and the long range electrostatic interactions were treated with the particle mesh Ewald method\cite{darden1993particle} with a real space cutoff of 0.9 nm. The Lennard-Jones potential was truncated at 0.9 nm. The LINCS algorithm\cite{hess1997lincs} was used to constrain bond lengths along with a 2 fs time step.

\subsection*{Order parameters}

We employ here two widely used order parameters for comparison: the tetrahedrality parameter and the local structure index. 

The tetrahedrality parameter is given by:

\begin{equation}
    q_i=1-\frac{3}{8}\sum_{j=1}^3\sum_{k=j+1}^4\left(\cos\theta_{jik}+\frac{1}{3}\right)^2
\end{equation}

\noindent where $\theta_{jik}$ is the angle between the two vectors connecting the central molecule, $i$, to two neighbors, $j$ and  $k$.\cite{yagasaki2019liquid,errington2001relationship} The value of $q_i$ is maximized when the surrounding four molecules are located in a regular tetrahedral arrangement.

The local structure index (LSI) is obtained as follows. 
The set of radial oxygen-oxygen distances $r_j$ corresponding to the $N$ neighboring molecules that are
within a cutoff distance of 3.7 Å from the reference molecule $i$
are ordered: $r_1 < r_2 < \cdot\cdot\cdot < r_j < r_ {j+1} \cdot\cdot\cdot < r_N < 3.7 < r_{N+1}$.
The local structure index (LSI) value I is then defined as the inhomogeneity in this distribution of radial distances, i.e.,

\begin{equation}
    I(i)=\frac{1}{N(i)}\sum_{j=1}^{N(i)}\left[ \Delta(j,i)-\bar\Delta(i)\right]^2
\end{equation}
where $\Delta(j,i)$= $r_{j+1}-r_j$ and $\bar\Delta(i)$ is the average over all neighbours $j$ of a molecule $i$ within a given cutoff. Hence, $I$ provides a quantitative measure of the fluctuations in the distance distribution surrounding a given water molecule within a sphere defined by a radius of 3.7 \AA.\cite{martelli2019unravelling,shiratani1996growth,shiratani1998molecular} Thus, the
index $I$ measures the extent to which a given water molecule is surrounded by well-defined first and second coordination shells.

\section*{Results and discussion}

We analyze four temperatures along the 1950 bars isobar, that crosses the liquid-liquid coexistence line (at around 175-178 K for TIP4P/2005): 170 K (LDL phase), 180 K (HDL phase, just above the coexistence line), 200 and 240 K (HDL phase).\cite{biddle16} 
At each temperature we analyze 100 MD simulation frames extracted from the corresponding MD simulation in the graph theory framework. The sampling of configurations can easly be increased but we observed that we can already obtain converged results already with 100 frames.
We consider the water oxygen atoms as the network nodes and we consider two nodes as connected by an edge if their distance is less or equal to 0.35 nm. For each network we then compute the degree $deg$ and the node total communicability $NTC$ (see Methods).

\begin{figure}
\centerline{
\resizebox*{8.00cm}{!}{\includegraphics{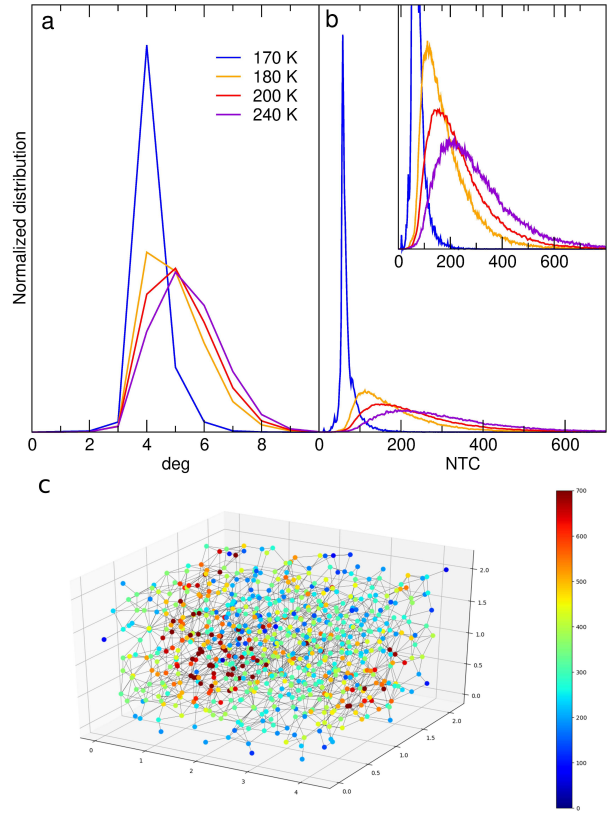}}}
\caption{Normalized distribution of the degree $deg$ (a) and node total communicability $NTC$ (b). Inset in panel b highlights the $NTC$ distributions in the HDL phase. (c) Topological distribution of the water nodes with different $NTC$ values for a representative structure at 240 K. It can be noted that the high $NTC$ values (from  dark red to green) are not uniformly  distributed but rather show a monotonic decrease with increasing distance from the regions with the highest values (dark red).}
\label{distro}
\end{figure}

\begin{figure}
\centerline{
\resizebox*{12.00cm}{!}{\includegraphics{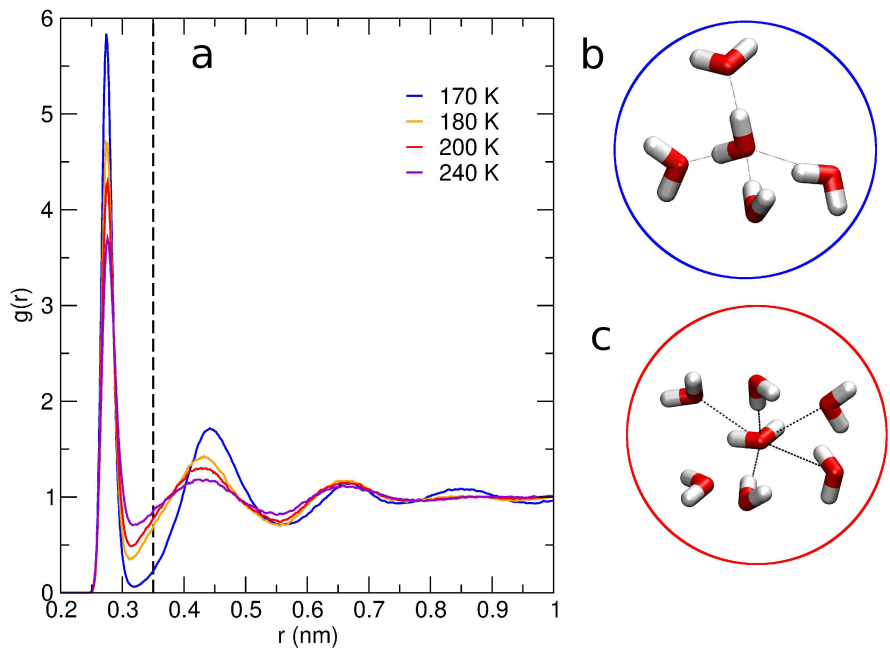}}}
\caption{a: Pair radial distribution functions ($g(r)$) for the O$\cdot\cdot\cdot$O contacts as obtained by the simulations at 1950 bars at 170 K (blue), 180 K (orange), 200 K (red) and 240 K (violet). b and c: Representative snapshots of the LDL tetrahedral arrangement (b) and of a HDL structure with interstitial water molecules at distance $d\le$0.35 nm (c).}
\label{rdf}
\end{figure}

The normalized distributions of the degree and the total communicability at 1950 bars (Figure \ref{distro}a and b) clearly show that both these centrality measures behave differently in the LDL phase with respect to the HDL one: in LDL both distributions are sharper and centered at lower values with respect to the corresponding ones in HDL. The degree, reporting on the average number of connections of each node, well agrees with what is expected for the LDL and HDL structures. In the low density regime almost all nodes feature 4 edges with a small population of nodes with 5 edges. In the high density regime there is a consistent population of nodes featuring 4, 5 and 6 edges. The balance between these three populations in HDL changes upon raising the temperature, consistently with a gradual increase of the packing around the first hydration shell. 
The differences observed at the various temperatures also well agree with what can be inferred from the temperature trend of the pair radial distribution functions $g(r)$ for the O$\cdot\cdot\cdot$O contacts (Figure \ref{rdf}). In particular, as expected, the radial cumulative distribution function provides a coordination number at the chosen cutoff (0.35 nm) that matches the average degree.  

The 0.35 nm cutoff is here chosen to include as connected nodes the interstitial water molecules that populate the space between the first and the second hydration shell. As a matter of fact, the first hydration shell was previously shown to be essentially the same in LDL and HDL and the importance of variations between the two phases in the second hydration shell, and in particular its shift towards lower radii upon raising the density, was pointed out.\cite{martelli2019unravelling,martelli2018local,soper2000structures,fanetti2014structure} For the sake of completeness, we also performed a test with two different additional cutoff lengths for the connections between the nodes: 0.32 nm and 0.37 nm, corresponding respectively to the first minimum of the radial distribution function and to the typical cutoff used for the calculation of the local structure index. As expected, with the 0.32 nm cutoff the differences between the LDL and HDL phase are less evident. With the 0.37 nm cutoff we do not observe relevant variations in the results.

While the degree provides the same information that can be obtained by computing the radial distribution function, we note that the latter is a global measurement that cannot be used to obtain local information. The degree, instead, can be computed for each water molecule at each time frame of the MD simulation, giving insights into the behavior of the system at the single molecule level. The degree is the simplest among the available centrality measures and its structural interpretation is straightforward, yet it has the drawback of being ``too local'', as it only takes into account the connectivity of the node under consideration.\cite{10.1093/comnet/cnt007} In addition, there is a relevant overlap between the degree distributions obtained in the LDL and HDL phases at 1950 bars, as both share a relevant number of nodes having 4 edges. These two drawbacks prevent the degree from being a fine enough parameter to distinguish between HDL and LDL. We also point out that the information that can be obtained by computing the degree could be also obtained, even at the single molecule level, with standard MD simulations post-processing tools which have been indeed frequently applied to investigate LDL and HDL water (e.g., by computing the $d_5$ parameter,\cite{cuthbertson2011mixturelike} see also below). 

We thus focus on the $NTC$, which has the same advantage of the degree (i.e., it can be computed as an instantaneous single-molecule property) but that features  distributions in the two phases that are much less overlapped (see Figure \ref{distro}b). 
In addition, the $NTC$ has the great advantage of taking into account also the connectivity of the neighbors of the node under consideration.
In fact, as shown in Eq. \ref{Eq:TC}, the $NTC$ is constructed by including all the possible walks (of all possible lengths) between the node under consideration and each other node in the graph, with a penalty factor for long walks. Therefore, the $NTC$ ranking of the $i$th node depends on its own connectivity (i.e, on the degree), and on the connectivity of its first, second, ..., neighbors (up to the farthest neighbors) with progressively decreasing weights. In the present case, we choose $\beta$=1 in Eq. \ref{Eq:TC}. With this choice, the weight of a walk of length $k$ is $1/k!$. 
Therefore, medium-long range effects are included in the ranking provided by the $NTC$. This allows us to take into account the packing of the whole hydration shell in the ranking of each water molecule. The choice of $\beta$=1 is the simplest possible choice that also implies a simple dependence on the length of the walks among the nodes. A deeper investigation of the sensitivity of the results on this parameter and the possibility of a tuning of $\beta$ to take into account temperature (and possibly pressure) effects will be addressed in future works.

Besides the evident difference in the $NTC$ distributions going from LDL to HDL (see Figure \ref{distro}b), we also note some variability in the distributions in the HDL phase (see inset of panel b, Figure \ref{distro}): upon increasing the temperature, the distribution is broader and the mode is shifted towards higher values.
This suggests that the $NTC$ variation can be used not only to distinguish between the low and high density regimes, but also to further characterize the HDL phase. 
 Observing the spatial distribution of the water molecules in the box, we note that there are regions with very high $NTC$ values. At increasing distance from these regions, the $NTC$ values decrease in a continuous way (see Figure \ref{distro}c). This suggests that in the HDL phase a continuum of states with different connectivity (and thus, local density) might be present. In particular, the regions at very high $NTC$ might be related to the presence of very-high density regions.

We define thus 3 $NTC$ regions: for $NTC \le 90$ we assign a water molecule to the LDL phase, with 90 being the $NTC$ value at the crossing point between the LDL and coldest HDL distributions (see Figure \ref{distro}b, blue and orange curves). The lowest temperature here considered at which water is in the HDL phase (180 K) is in fact just above the critical temperature at 1950 bars and therefore its distribution is representative of the coldest HDL at this pressure. 
While we recognize that the $NTC$ distributions in the HDL phase do not provide a clear indication of the existence of two subpopulations, we believe that the asymmetric non-gaussian tail at high values of the $NTC$ distributions could be representative of the presence of VHDL molecules. Therefore, we use the two following regions.
For $90 < NTC \le 350$ we assign a water molecule to the HDL region. This region corresponds to the most populated $NTC$ zone at all the investigated temperatures at which water is expected in the HDL phase. 
For $NTC > 350$ (roughly corresponding to the 84\% of the $NTC$ distribution at 200 K, i.e. +$\sigma$ for a gaussian distribution) we assign a water molecule to the VHDL region, characterized by a very high connectivity in terms of total communicability and with nodes packed in a dense and highly connected network.

\begin{figure}
\centerline{
\resizebox*{8.00cm}{!}{\includegraphics{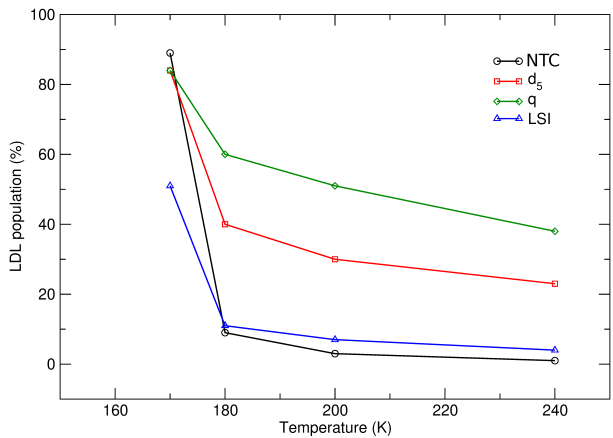}}}
\caption{LDL population as a function of the temperature as obtained by defining the LDL phase according to the $NTC$ values (black circles), the $d_5$ parameter (red squares), the tetrahedrality parameter (green diamonds) and the local structure index (blue triangles).}
\label{pop:all}
\end{figure}

On the basis of the above definition, we analyze the LDL fraction at the four temperatures here investigated. As it can be observed in Figure \ref{pop:all}, the $NTC$ very well differentiates the two phases, being able to assign a LDL population of $\approx$90\% at 170 K and a HDL population from $\approx$90\% to $\approx$99\% at the three higher temperatures. While small LDL clusters can be present in the HDL phase (and $vice versa$),\cite{martelli2019unravelling} the very high population of LDL at 170 K and of HDL at the other three temperatures is consistent with a transition between the two pure phases, as expected along the 1950 bars isobar that crosses the coexistence line. 

As already mentioned, a number of parameters is commonly used to assign water molecules to the LDL or HDL phase along a MD simulation.\cite{cuthbertson2011mixturelike,shiratani1996growth,shiratani1998molecular,biddle16,errington2001relationship} Therefore, we compare the above obtained populations with the corresponding populations provided by three of these well-known parameters, namely the parameter $d_5$, the tetrahedrality parameter $q$ and the local structure index (LSI, also denoted by $I$). The $d_5$ parameter is a very simple order parameter based on the distance $d_5$ to the fifth nearest neighbor: a molecule is assigned to LDL when the distance $d_5$ is greater than the cutoff distance $r_0$ = 0.35 nm.\cite{biddle16} The definition of the tetrahedrality parameter $q$ and the local structure index are provided in the Methods section. The results obtained with these three parameters on the present four MD simulations are schematized in Figure \ref{pop:all} and compared to the ones obtained by using the $NTC$.

As previously recognized,\cite{biddle16} the $d_5$ parameter significantly underestimates the HDL component in the pure HDL phase, as it assigns to the LDL phase all the molecules having four neighbors at a distance smaller than the cutoff distance. However, the degree (Figure \ref{distro}a), with the same 0.35 nm cutoff, shows that a relevant number of water molecules in the HDL phase features 4 edges even at relatively high temperatures.
The tetrahedrality parameter $q$ is based on the specific structural properties of LDL (see Methods): the value of $q$ is maximized when the surrounding four molecules are located in the regular tetrahedral arrangement typical of the low density phase.\cite{yagasaki2019liquid} The distribution of $q$ as obtained from the present MD simulations at the four temperatures is reported in Figure 1 of the Supporting Information (SI) and shows that the four curves are significantly overlapped. Because of this, it is not straightforward to identify an absolute criterion to assign a molecule to the LDL or the HDL phase. We thus borrowed the criterion used by Tanaka and coworkers,\cite{yagasaki2019liquid} who defined that a water molecule is a low-$q$ molecule when its $q$ value is lower than that averaged over all molecules at each time step. We define therefore a molecule as belonging to the LDL phase if its $q$ value is higher than that averaged over all the molecules and time frames of the four MD simulations.

Due to the already mentioned overlap in the distributions at the different temperatures, the HDL component is severely underestimated by $q$ in the HDL phase, reaching a 60\% population at 240 K.
To evaluate the fraction of LDL and HDL with the LSI, we use the same criterion used by Martelli in different conditions of temperature and pressure\cite{martelli2019unravelling} and the same definition of the isosbestic point ($I_{is}$=0.13 \AA$^2$). We note that, while the LSI catches a very high fraction of HDL molecules in the pure HDL phase at high temperatures, it severly underestimates the LDL fraction in the pure LDL phase. The reason for this can be again observed in the distribution of LSI (see Figure 2 in the SI). Upon raising the temperature, there is an increase in the population of water molecules with low LSI values (i.e., molecules in locally disordered environments) and a decrease in the population of water molecules with higher LSI values (i.e., molecules in locally ordered environments). However, at 170 K the low LSI component is still very relevant, leading to an overestimation of the HDL fraction.

Overall, the present data suggest that the total communicability performs very well in distinguishing between the LDL and HDL phases, also when compared to more standard and widely used parameters. We also point out that, to obtain the network properties emerging from the present analyses, no structural properties are hypothesized for any of the two phases, as done for example by the tetrahedrality parameter or by the $d_5$ parameter, that assumes {\it a priori} the presence of interstitial water molecules between the first and second hydration shell in the HDL phase. 

In this view, it is particularly interesting to investigate the topological features of the three regions (LDL, HDL and VHDL), to characterize the internal structural arrangement of water at different temperatures.

\begin{figure}
\centerline{
\resizebox*{18.00cm}{!}{\includegraphics{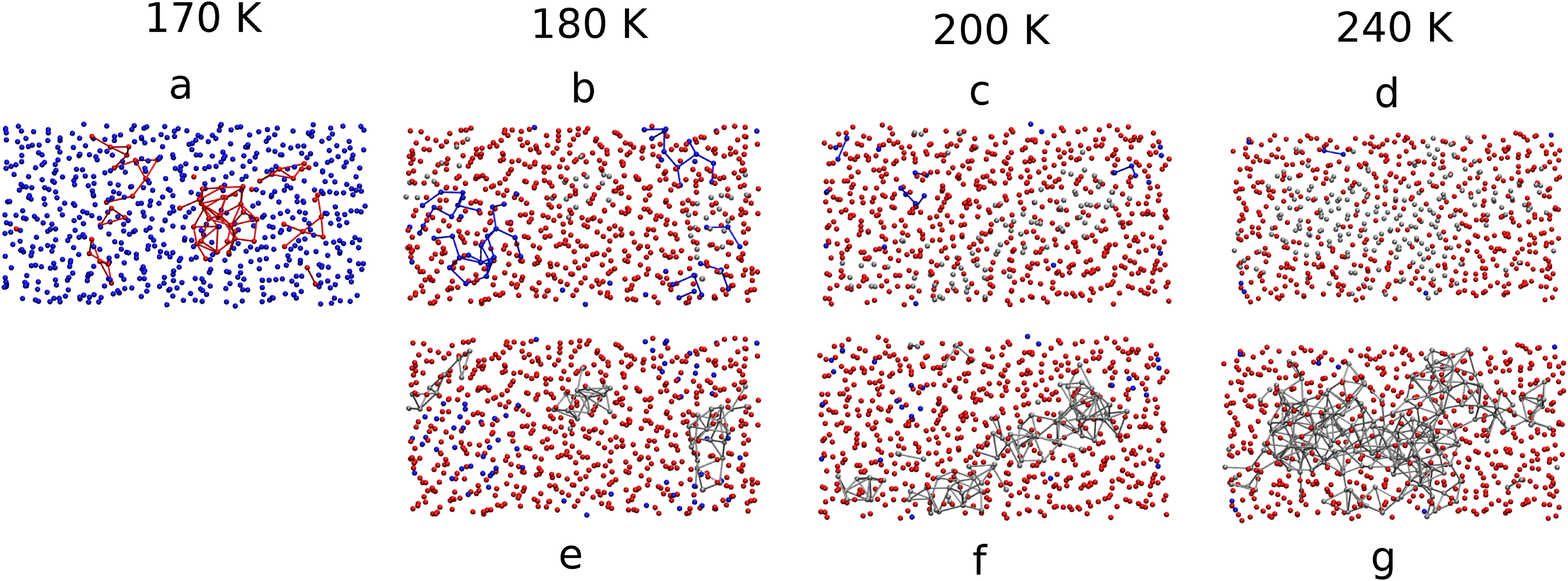}}}
\caption{Representative snapshots of the topological arrangement of the four $NTC$ regions at 170 K (a), 180 K (b,e), 200 K (c,e) and 240 K (d,g). Blue nodes represent LDL molecules, 
red nodes HDL molecules and silver nodes VHDL molecules. In panel a, red edges highlight the connections among the HDL molecules. In panels b to d, blue edges highlight the connections among the LDL molecules. In panels e to f, silver edges highlight the connections among the VHDL nodes.} 
\label{box}
\end{figure}

At 170 K, the HDL component (11\%) is mainly assembled into small clusters: isolated 
HDL (red) nodes are only rarely present (see Figure \ref{box}a).
At 180 K we observe a somehow specular behavior: the small ($\approx$9\%) LDL component (blue) is organized in small clusters surrounded by the HDL molecules. However, LDL molecules have a lower tendency to form well packed clusters, and isolated LDL nodes can be more frequently observed (see Figure \ref{box}b). 
At 200 K and 240 K a small amount of LDL is still present ($\approx$3\% and $\approx$1\% respectively). This low population does not allow the formation of extended LDL clusters, leaving the LDL nodes mostly isolated (see Figure \ref{box}c and d).

While at 170 K, i.e. in the LDL phase, no VHDL molecules are detected, at 180 K (i.e. just above the critical temperature),  a non negligible population of VHDL nodes (6\%), that are almost all clustered in packed assemblies and surrounded by HDL molecules, is observed (see Figure \ref{box}e). As already mentioned, 180 K is the coldest HDL at this pressure. Therefore, the presence of clusters characterized by a very high connectivity seems to be an intrinsic feature of the HDL phase under pressure, even at low temperatures.
At 200 K the population of VHDL increases (14\%), with the formation of more extended clusters (Figure \ref{box}f).
At 240 K, where the population of VHDL nodes is further increased (28\%), big VHDL clusters can be observed (see Figure \ref{box}g). 
As already anticipated, an interesting feature of these large clusters is that a specific spatial distribution of the {\em NTC} values is observed within each cluster, with the highest values being located in the central part of the cluster and with decreasing values at increasing distances from the center (see Figure \ref{distro}c for a representative example).

\begin{figure}[ht!]
\centerline{
\resizebox*{14.50cm}{!}{\includegraphics{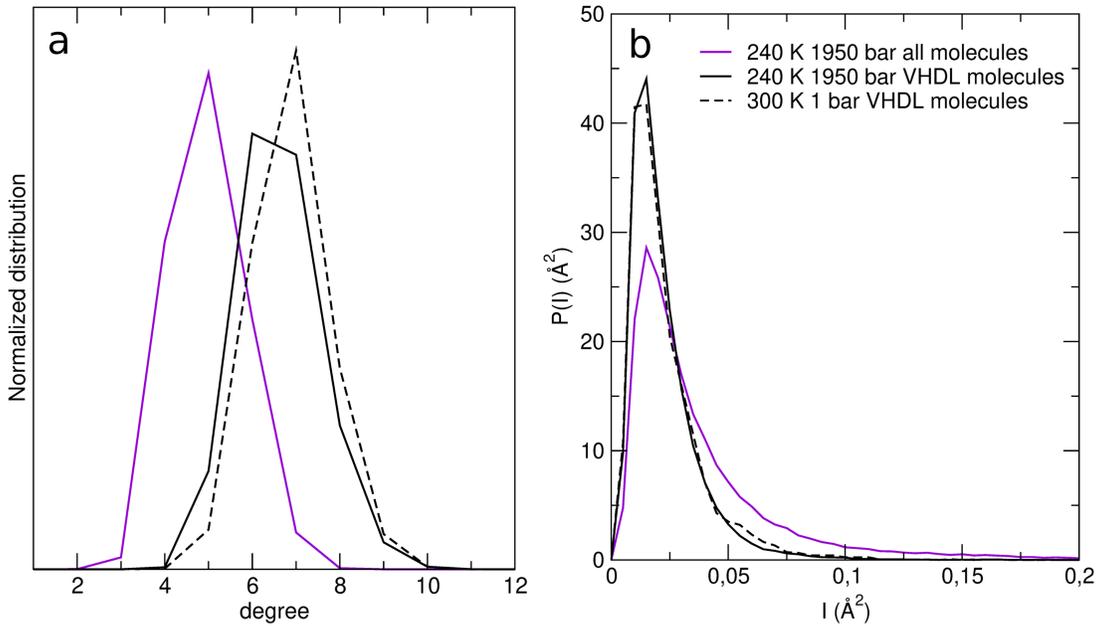}}}
\caption{Degree, $deg$, distribution (a) and LSI distribution (b) computed at 240 K and 1950 bars for all the molecules in the simulation box (violet), at 240 K and 1950 bars for the VHDL molecules only (solid black) and at 300 K and 1 bar for the VHDL molecules only (dashed black).}
\label{rdfVHTC}
\end{figure}

To better characterize the VHDL clusters, we compute the degree 
on the subset of VHDL molecules at 240 K. 
Interestingly, this analysis (Figure \ref{rdfVHTC}a) shows the presence of at least two interstitial water molecules within a 0.35 nm cutoff (corresponding to a degree of 6), in agreement with what hypothesized for the very high density amorphous.\cite{fuentes15}
Remarkably, while evidences of the presence of VHDL were previously obtained only by increasing the total density of the simulated system, we show here that VHDL clusters are present in pure HDL and are thus an intrinsic feature of this phase, at least under moderate pressure.
To understand if VHDL clusters are also present at ambient conditions, we perform the same analysis on 100 frames of a 50-ns long MD simulation at 300 K and 1 bar. By means of the same $NTC$-based definitions used at 1950 bars, we obtain
at ambient conditions a 0.08 fraction of VHDL molecules. The degree computed on this subset of VHDL molecules (Figure \ref{rdfVHTC}a) again shows the presence of at least two interstitial water molecules within a 0.35 nm cutoff. 

Notably, the LSI distributions calculated on the subpopulation of VHDL molecules identified with the $NTC$ are only slightly different from those calculated on all the molecules at 240 K (Figure \ref{rdfVHTC}b). As already recognized, the LSI parameter well describes the HDL phase and is sensitive to the fluctuations of the second shell. Nonetheless, 
the LSI distributions of the VHDL molecules are only slightly sharper than the global ones, lacking the tail at higher values.  
Overall, the two VHDL distributions are very similar to the global one and the global LSI curve at 240 K does not allow identification of the VHDL subpopulation. 
Identification of the VHDL clusters is instead possible by using the $NTC$, that appears a promising descriptor of the structural properties of fluids at the microscopic level. While the degree is very useful to detect the presence of interstitial water molecules, the $NTC$, taking into account the cumulative connectivity of each node, helps in identifying larger structures with a long range organization like clusters.

\section*{Conclusions}

We characterize here the structure of liquid water at the microscopic level using molecular dynamics simulations and the calculation of centrality measures derived from graph theory. In particular, we use the node total communicability to investigate the evolution of the high and low density phases (HDL and LDL) along the 1950 isobar, that crosses the coexistence line. The node total communicability discriminates rather well the two liquid phases differing in density and, thanks to that, we are able to characterize their structural arrangement at different temperatures. Interestingly, we also identify clusters of the very high density form of liquid water (VHDL), both at 1950 bars and at ambient conditions (1 bar and 300 K). 
Identification of these cluster was possible thanks to the use of the node total communicability as a descriptor of the structural properties at the microscopic level. This descriptor reports on the cumulative connectivity of each water molecule and highlights the contribution of long-range effects. A further improvement of the structural description of the liquid phase that we will address in the next future is the representation of the water network as an oriented graph in which the connections will be defined to explicitly take into account hydrogen bonds among water molecules. Nonetheless, we show that this graph-theory based approach is able to well distinguish between HDL and LDL even if protons are not explicitly included. We also show that, as already suggested, inclusion of long range many-body effects is crucial for the description of disordered structures.

\section*{Supporting Information}

Distribution of $q$ and LSI at all temperatures.

\bibliography{references_chiara}
\bibliographystyle{unsrtnat}

\end{document}

\maketitle

\begin{figure}
    \centering
    \includegraphics[width=12cm]{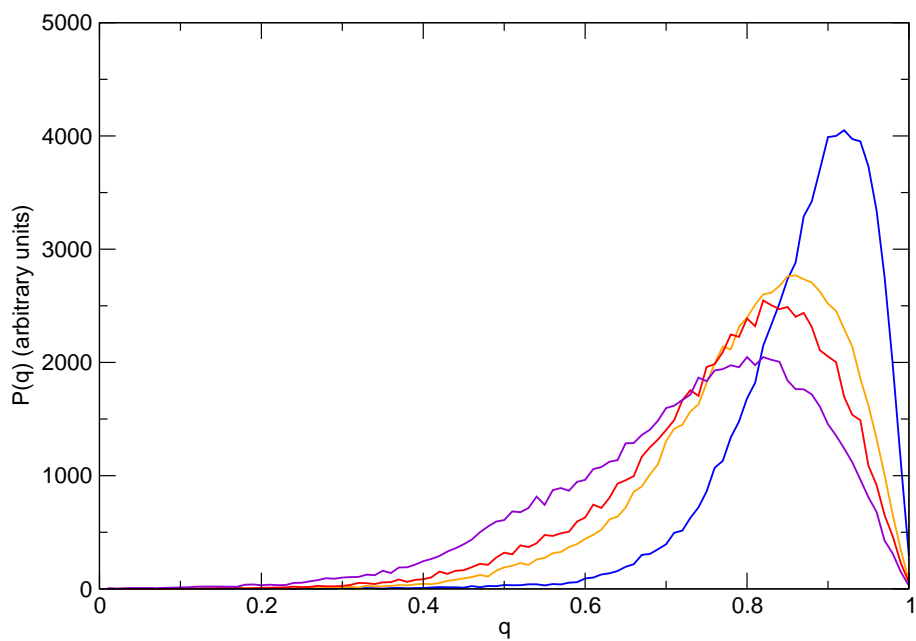}
    \caption{Distribution of the tetrahedrality parameter $q$ as obtained from the MD simulations at 1950 bars and 170 K (blue), 180 K (orange), 200 K (red), 240 K (violet).}
    \label{qdist}
\end{figure}

\begin{figure}
    \centering
    \includegraphics[width=12cm]{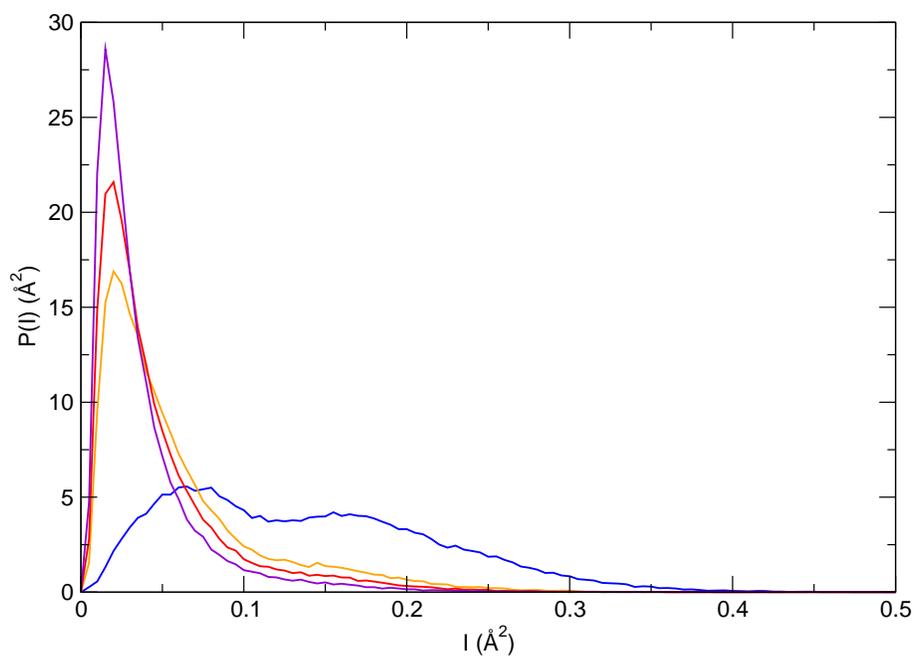}
    \caption{Distribution of the local structure index, LSI, as obtained from the MD simulations at 1950 bars and 170 K (blue), 180 K (orange), 200 K (red), 240 K (violet).}
    \label{fig:my_label}
\end{figure}